# Revisiting the science case for near-UV spectroscopy with the VLT

C. J. Evans[1], B. Barbuy[2], B. Castilho[3], R. Smiljanic[4], J. Melendez[2], J. Japelj[5],
S. Cristiani[6], C. Snodgrass[7], P. Bonifacio[8], M. Puech[8], A. Quirrenbach[9]

[1] UK Astronomy Technology Centre, Royal Observatory, Blackford Hill, Edinburgh, EH9 3HJ, UK
[2] Universidade de São Paulo, IAG, Rua do Matão 1226, Cidade Universitária, São Paulo, 05508-900, Brazil
[3] Laboratório Nacional de Astrofísica/MCTIC, Rua Estados Unidos, 154 - 37504-364, Itajubá, MG, Brazil
[4] Nicolaus Copernicus Astronomical Center, Polish Academy of Sciences, Bartycka 18, 00-716, Warsaw, Poland
[5] Astronomical Institute Anton Pannekoek, University of Amsterdam, Science Park 904, 1098 XH, Amsterdam, the Netherlands
[6] INAF - Osservatorio Astronomico di Trieste, via G. B. Tiepolo 11, 34131 Trieste, Italy
[7] School of Physical Sciences, The Open University, Milton Keynes, MK7 6AA, UK
[8] GEPI, Observatoire de Paris, PSL University, CNRS, 5 Place Jules Janssen, 92190 Meudon, France
[9] Landessternwarte, Zentrum für Astronomie der Universität Heidelberg, Königstuhl 12, 69117, Heidelberg, Germany

**ABSTRACT**

In the era of Extremely Large Telescopes, the current generation of 8-10m facilities are likely to remain competitive at far-blue visible wavelengths for the foreseeable future. High-efficiency (>20%) observations of the ground UV (300-400 nm) at medium resolving power ($R$~20,000) are required to address a number of exciting topics in stellar astrophysics, while also providing new insights in extragalactic science. Anticipating strong demand to better exploit this diagnostic-rich wavelength region, we revisit the science case and instrument requirements previously assembled for the CUBES concept for the Very Large Telescope.

## 1. INTRODUCTION

The European Extremely Large Telescope (ELT) is now under construction in northern Chile by the European Southern Observatory (ESO). With a primary aperture of 39m, the ELT will be unprecedented in its light-gathering power, coupled with exquisite angular resolution via correction for atmospheric turbulence by adaptive optics (AO). In contrast to current large telescopes such as ESO's Very Large Telescope (VLT), AO is an integral part of the ELT, which has a novel five-mirror design including a large adaptive mirror (M4) and a fast tip-tilt mirror (M5). The choice of protected silver (Ag+Al) for the ELT mirror coatings (excl. M4) ensures a durable, proven surface with excellent performance across a wide wavelength range, but the performance drops significantly in the blue-visible part of the spectrum compared to bare aluminium. ESO are actively researching alternative coatings, but in the short-medium term we can assume that the performance of the ELT in the blue-visible will be limited in this regard. Indeed, during the Phase A study of the MOSAIC multi-object spectrograph, Evans et al. (2016) concluded that a blue-optimised instrument on the VLT could potentially be competitive with the ELT at wavelengths shorter than 400 nm, and this was adopted for the blueward limit of the study (later revised to 450 nm from similar arguments re: end-to-end efficiency).

Motivated by this, we have revisited the Phase A study undertaken in 2012 of the Cassegrain U-band Brazilian-ESO Spectrograph (CUBES). The CUBES study investigated a spectrograph operating at 'ground UV' wavelengths (spanning 300-400 nm) to open-up exciting new scientific opportunities compared to the (then) planned instrumentation suite for Paranal Observatory. Overviews of the CUBES science case and Phase A concept for a high-throughput, medium resolution ($R$~20,000) spectrograph were presented by Barbuy et al. (2014) and Bristow et al. (2014).

The cases assembled for the CUBES study required medium-resolution ($R$~20,000) spectroscopy, with the design focusing on maximizing the instrument throughput to best exploit the VLT in this wavelength domain (e.g. ensuring high signal-to-noise, S/N > 50, observations of solar-type stars down to $V$~16 mag in a one-hour observing block). Both UVES (Dekker et al. 2000) and X-Shooter (Vernet et al. 2011) provide spectral coverage down to the blue atmospheric cut-off at Paranal (shortwards of 320 nm) but UVES only has an efficiency of <5% shortwards of 380 nm (meaning many targets are simply out of reach) while X-Shooter is limited to $R$~6,000 in the blue arm. We note that the arrival of ESPRESSO (Pepe et al., these proceedings) at the VLT will present fantastic opportunities for high-resolution visible spectroscopy, but its wavelength coverage only extends as far bluewards as 380 nm. As such, the CUBES concept remains compelling for the future of the VLT, and will provide unique capabilities well into the era of ELT operations.

Here we refresh the science case for such a ground-UV spectrograph on the VLT to take stock of developments since the Phase A study, and to revisit the instrument requirements before planning technical work to advance the concept further. Although motivated by the cases discussed below, we also envisage this instrument as a chance to open-up new parameter space at Paranal, with the expectation of a broader range of scientific applications than those outlined here, and with a diverse user community. The Galactic and extragalactic science cases are described in Sections 2 and 3, respectively, followed by some concluding remarks and next steps in Section 4.

## 2. GALACTIC SCIENCE

The general Galactic theme is the study of elemental abundances which are inaccessible with current instruments due to insufficient sensitivity and/or spectral resolution. The cases summarised below are primarily motivated by efforts to characterise the histories of different stellar populations of the Milky Way, with a separate case in planetary science introduced in Section 2.4.

### 2.1. Heavy elements in metal-poor stars

Ongoing large-area surveys are discovering increasing numbers of old, metal-poor Galactic stars (e.g. SkyMapper, Keller et al. 2014, Jacobson et al. 2015; Pristine, Starkenburg et al. 2017). Such stars provide an excellent means to explore the production of heavy elements via supernovae (SNe) and stellar nucleosynthesis at early times in the Galaxy. A key question is still whether low abundances of what had assumed to be slow-neutron-capture ('*s*-process') elements are actually the products of rapid-neutron-capture ('*r*-process') nucleosynthesis (e.g. Truran, 1981) or if they are from genuine *s*-process channels such as weak *s*-process in massive stars, in stars on the asymptotic giant branch (AGB), or from rapidly-rotating massive stars (e.g. Frischknecht et al. 2016).

Moreover, the formation channels for the *r*-process are particularly topical, with the suggestion of *r*-process elements in the ejecta associated with the GW170817 kilonova explosion from a binary neutron-star merger (Pian et al. 2017; Smartt et al. 2017). This nucleosynthesis is thought to occur both during the merging and in the milliseconds afterwards. Other predicted sites of *r*-process nucleosynthesis include: 1) neutrino-driven winds arising from core-collapse SNe, where sophisticated relativistic models predict nucleosynthesis over the full mass range ($90<A<240$, e.g., Wanajo et al. 2014; Goriely et al. 2015); 2) in electron-capture SNe, which might account for the weak *r*-process or the lighter-element primary process (e.g. Bisterzo et al. 2017); 3) magnetohydrodynamically-driven jets from core-collapse SNe, resulting from massive stars characterized by a high rotation rate and a large magnetic field (Nishimura et al. 2015).

Spectral lines from heavy elements are mostly located in the near-UV because of atomic physics, and for many elements the lines in this region are the only spectral information we have. These include strong lines of first-peak elements ($38<Z<48$) YII, ZrII, NbII, PdI, AgI; second-peak elements ($56<Z<72$) BaII, LaII, CeII, NdII, EuII, GdII, TbII, DyII, HoII, ErII, TmII; third-peak elements ($76<Z<88$) OsI, IrI, PbI, BiI; and the actinides ($Z>89$) ThII, UII.

### 2.2. CNO abundances from molecular lines

As the most common (metallic) elements in the Universe, abundances of C, N, and O are key inputs to studies of both stellar evolution and the chemical evolution of galaxies. There are absorption lines of CN, NH and OH in the 300-400 nm region which provide important diagnostics for estimates of CNO abundances (see main panel of Fig. 1). For instance, photometric observations with the *Hubble Space Telescope (HST)* over 250-450 nm are sufficiently sensitive to the relative strengths of these bands to begin to decipher the multiple stellar populations of globular clusters (e.g. Piotto et al. 2015), but higher-resolution spectroscopic follow-up is required to provide precise estimates of abundances.

In particular, the ground-UV molecular lines can provide greater precision for CNO abundance estimates in metal-poor stars than relying on diagnostics at longer wavelengths. For instance, O abundances estimated using the A–X OH transitions in this region were in better agreement with analysis of the [O I] 630.03 nm forbidden line than the O I triplet at 777.1-5 nm (García-Pérez et al. 2006). Although the strongest OH features lie bluewards of the atmospheric cut-off, there are several OH lines accessible longwards of 300 nm, as illustrated by, e.g., spectral fits to OH lines from UVES spectra of metal-poor subgiants (see Fig. 9 of García-Pérez et al. 2006).

As we attempt to better constrain the early chemical evolution of the Galaxy, the O abundances derived from the UV OH lines are inconsistent with OH lines in the near-IR as well as results in the visible. To overcome the limitations of 1D LTE analyses, recent efforts have focused on 3D hydrodynamical models (e.g. Prakapavičius et al. 2017), which find better agreement with IR diagnostics, but we remain starved of observations with which to test the latest models.

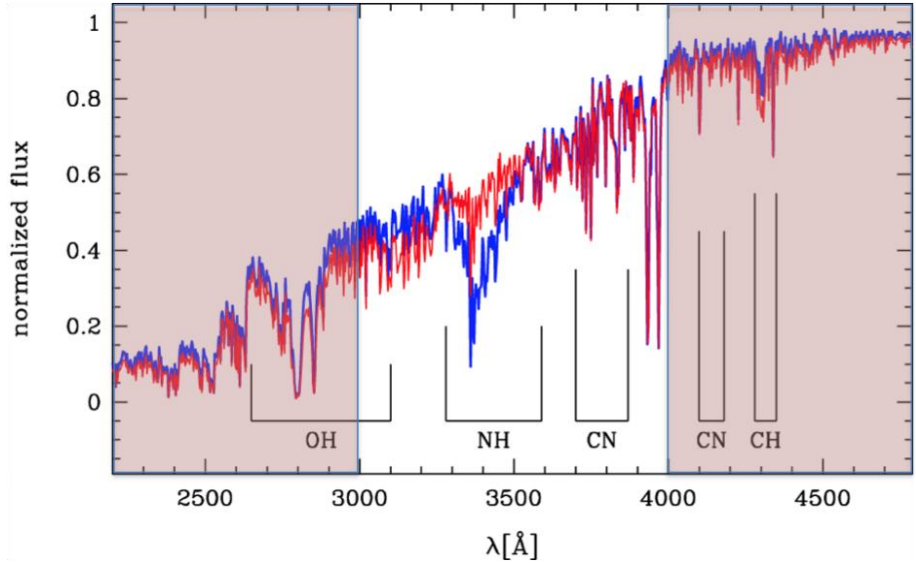

Fig. 1: Absorption features from OH, NH and CN in the 300-400 nm window for N-rich (blue) and N-poor (red) simulated spectra of stars on the red-giant branch (adapted from Fig. 1 of Piotto et al. 2015).

### 2.3. Beryllium: New insights into stellar atmospheres and Galactic chemical evolution

Beryllium abundances are a powerful diagnostic across a broad range of stellar science. The ground UV is our only opportunity to observe spectral lines of Be, with two Be II resonance lines at 313.042 and 313.107 nm. Four cases pertaining to Be were developed in the CUBES Phase A study - these remain both relevant and unique and are summarised in the following sections. We also include a short new case related to possible Li enrichment via the decay of $^7$Be in novae.

#### 2.3.1. Using solar twins to constrain Li and Be depletion

For many years the Sun has been the cornerstone for stellar astrophysics, but we can now reverse this by using observations of solar twins to further our understanding of the physical processes that occur in the solar interior. Solar twins are stars that only differ slightly from the Sun in mass, age, rotation rate and metallicity, such that the general principles used for solar models are still valid[1].

The depletion of Li is a key test of solar models, and depletions of Be provide an important, independent test of the mechanisms involved (e.g., mass, age, treatment of convection, additional mixing, mass loss, metallicity). Be is destroyed at higher temperatures than Li ($3.5\times10^6$ cf. $2.5\times10^6$ K) so its expected depletion is much smaller, with observations from Boesgaard & Krugler (2009) showing that its relative depletion is about one third that of Li (thus high S/N data is required). Their observations of 'solar-mass' stars with Keck-HIRES suggested the Sun as Be-rich. However, their sample stars all had $M_V < 4.5$ compared with $M_V = 4.81$ for the Sun, suggesting their stars were more massive than the Sun and that they may not be good solar analogues. Nevertheless, their result is quite puzzling: how could solar-analogues have less Be and more Li than the Sun? If their Be abundances are lower, then Li should be even more depleted given that it is easier to destroy.

---

[1] This assumption breaks down quickly for solar analogues with masses > ~$1.1 M_\odot$ because they develop a convective core, radically changing the description of the stellar interior.

The first homogeneous study of Be in solar twins and the Sun, shows that Be is roughly constant with age (Tucci Maia et al. 2015). This means that the processes that deplete Li, are not deep enough as to also deplete Be. However, these results are from only a few solar twins and we need a much larger sample of Sun-like stars (several stars per Gyr bin, so ~50 in total) to investigate possible variations of Be with age. Interestingly, some anomalous solar analogs with low Be abundances were discovered by Takeda et al. (2011), and they seem related to stars with white dwarf companions (Desidera et al. 2016). A similar case has been found by Schirbel et al. (2015), where a solar twin depleted in Be seems to have been contaminated with neutron-capture elements by a former AGB companion, now likely a white dwarf.

### 2.3.2. Mixing and stellar physics

Moving beyond attempts to improve the solar model, Li abundances can also potentially tell us about mixing between the surface and interior layers of low-mass stars. Standard models of stellar evolution, which only include mixing in the convective layers, fail to explain nearly all of the observed patterns of surface Li abundances in low-mass stars (e.g., Boesgaard & Tripicco, 1986; Soderblom et al. 1993; Pasquini et al. 1997; Lebre et al. 1999; Sestito & Randich, 2005; Bouvier et al. 2016; Lyubimbov, 2016; Anthony-Twarog et al. 2018). These observations suggest physical processes beyond convection, with atomic diffusion, mass loss, magnetic fields, rotationally-induced mixing and mixing by internal gravity waves each proposed to explain the observations (and combinations of these processes). As noted in Section 2.3.1, Li and Be burn at different temperatures (corresponding to different depths in the stellar interiors). Abundances of Be can therefore provide a valuable complement to estimates for Li, helping to constrain the nature and description of the transport mechanisms of chemical mixing and angular momentum inside low-mass stars (Smiljanic et al. 2010, 2011). We have very sparse information on Be abundances given the limitations of current instruments – only a spectrograph optimised for the ground UV on an 8-10m class telescope can open-up the observations required.

### 2.3.3. Beryillium as a cosmochronometer

The stable isotope of beryllium ($^9$Be) is a product of cosmic-ray spallation of heavy (mostly CNO) nuclei in the interstellar medium in the early Galaxy and behaves as a primary element (Reeves et al. 1970). Its production is expected to be relatively uniform (assuming cosmic rays travel freely across the Galaxy), with relatively homogeneous abundances at a given time in the early life of the Milky Way (cf. the larger range expected from the products of stellar nucleosynthesis). Thus, Be abundances should correlate well with time and, compared to oxygen (or alpha-element) abundance, could provide a chronometer to investigate the evolution of the star-formation rate. In attempting this, it is important to observe main-sequence stars with undepleted Li, i.e. that did also not deplete their initial Be abundance.

An example from UVES observations is shown in Fig. 2 (from Smiljanic et al. 2009), in which stars in the thick disk appear to correlate with increasing α with time (i.e. to smaller Be abundances as it is only destroyed after primary production). In contrast, the halo stars appear to separate into two different sequences, interpreted as a difference in the star-formation histories of the two groups. The interpretation of these results is still open (e.g. they may be related to accretion of external systems or to different star-formation rates in initially independent parts of the early halo), but they demonstrate that the halo is not a uniform population with a single age-metallicity relation. A similar division was found by Nissen & Schuster (1997, 2010) using Fe as a tracer of time, but the division between the groups is clearer when using Be. In short, studies of Be abundances can potentially improve our knowledge of early Galaxy formation (e.g. Tan & Zhao, 2011), but larger observational samples of (kinematically selected) metal-poor stars are required to provide robust constraints on theoretical efforts.

There is also interest in Be abundances in the context of extremely metal-poor stars, where early results with UVES by Primas et al. (2000a, 2000b) suggested a flattening at [Fe/H] ~ −3 in the relation of log(Be/H) vs. [Fe/H]. There are a number of interesting scenarios that could produce such a plateau in the Be abundance, such as:
- Significant production of Be by inhomogeneous primordial nucleosynthesis;
- Pre-Galactic production by cosmic-rays in the intergalactic medium;
- Accretion of metal-enriched interstellar gas onto metal-poor halo stars when crossing the Galactic plane.

Unfortunately, there are few stars with estimates for Be in this regime and the notion of a plateau has been challenged (e.g. Boesgaard et al. 2011). Other stars at such low metallicities are too faint to be observed with current instruments, so only with increased sensitivity can we test for the presence of such a flattening (although this will remain challenging given the very low equivalent widths in such stars).

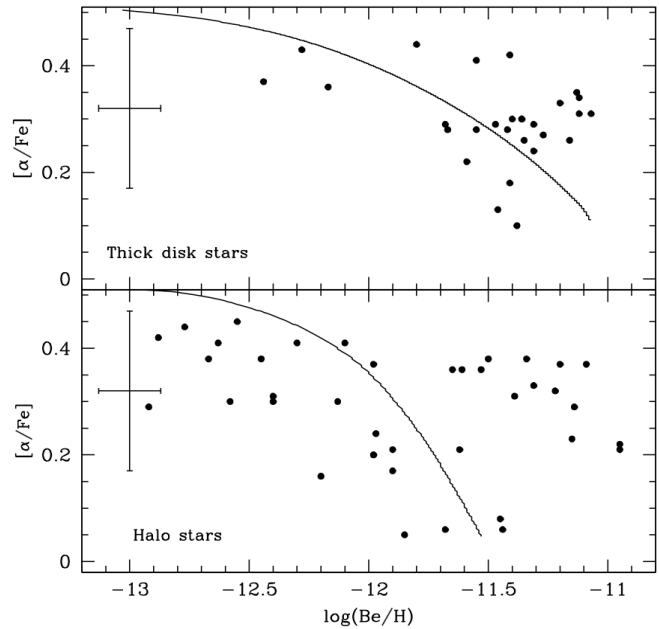

Fig. 2: Using Be abundances to investigate the early star-formation history of the Galaxy. Results for [α/Fe] and log(Be/H) from Smiljanic et al. (2009) for samples of stars in the halo and thick disk, compared to predictions of the evolution of Be with time (solid line, from Valle et al. 2002); there appear to be two distinct populations in the halo, suggesting different star-formation histories and/or origins.

### 2.3.4. Clues to the formation of globular clusters

Debate continues regarding the origins of the chemical inhomogeneities of light elements (Li, C, N, O, Na, Mg, Al) in globular clusters, which manifest as signatures of material processed by proton-capture reactions. These abundance patterns are thought to arise from a previous generation of stars polluting those observed now, but the source of this pollution remains unclear (see Bastian & Lardo, 2018, and references therein).

The combination of Li and Be abundances is again potentially very powerful. Both elements are destroyed by (p,α) reactions but Li is thought to be produced in AGB stars (via the Cameron-Fowler mechanism), while there is no production mechanism for additional Be. Other proton reactions (CNO and NeNa cycles) require much higher temperatures than the $3.5\times10^6$ K threshold for destruction of Be, so any further processed material will be devoid of Be. Pushing the capabilities of UVES, Pasquini et al. (2004) observed two turn-off stars in NGC 6397, finding very different O abundances but possibly the same Be abundance. This is a potentially profound result as the Be abundance should always be diluted in O-poor stars as it should only be present in the primordial material. This would present a significant challenge to current theories of the early evolution of globulars if confirmed, but to test this result we require significantly improved sensitivity cf. UVES to obtain Be abundances in a range of systems for the first time.

### 2.3.5. $^7$Be decay in novae as a potential channel for Li production

The large majority of Li is thought to be produced in the Big Bang and, as discussed above, is then destroyed in stars as they evolve. However, the overall abundance of Li is seen to grow with increasing stellar metallicity, at least up to solar [Fe/H] (Guiglion et al. 2016; Fu et al. 2018), and the source of Li is still an open question. Potential channels for Li production include cosmic rays, evolved low-mass stars, novae, and SNe explosions. For example, hot-bottom-burning in red giants is a candidate for Li production, with Be abundances providing an essential tracer of the process (Castilho et al. 1999). Recent observations of novae with Subaru-HDS have detected $^7$Be, that can decay to $^7$Li, providing a possible source of Li enrichment (Tajitsu et al. 2015). The efficiency of HDS in the ground-UV is slightly better than UVES (e.g. Fig. 7 from Noguchi et al. 2002), but a more efficient spectrograph on the VLT will be a particularly useful tool in this rapidly-developing topic.

## 2.4. Searching for water in the asteroid belt

The search for water in our solar system is far from complete and is tremendously difficult to undertake from ground-based facilities given the large water content of Earth's atmosphere. Detecting ice on the surface of distant bodies can be achieved with infrared spectroscopy, but only for the largest/brightest objects. For smaller bodies the ice is sub-surface and we must look for outgassing water escaping into space, e.g. the characteristic coma and tails of comets. The most powerful probe of this outgassing water from the ground is the hydroxide (OH) emission at 308 nm in the near-UV (e.g. Fig. 3). While observations in the far-UV or IR from space would be more sensitive, they would also be *substantially* more expensive and ground-based OH observations are the most compelling next step (see Snodgrass et al. 2017a).

The OH line is only currently observable for a few active comets while near the Sun and Earth. This severely limits studies of water production in comets around their orbits and we miss the seasonal effects that the *Rosetta* mission has recently revealed to be important (e.g. Kramer et al. 2017). It also prevents study of the vast majority of comets which are simply too faint. Even more tantalysing are studies of main-belt comets – bodies in asteroidal orbits that are seen to undergo activity (usually detected by a dust tail or trail) which is thought to arise from sublimation (e.g. recurrent activity near perihelion). They have typical sizes that are very common in the asteroid belt, so detection of outgassing water would point to a potentially large population of icy bodies, hence a large reservoir of water, of considerable interest in the context of models of the formation and evolution of the inner solar system (e.g. O'Brien et al. 2018).

This case provides a strong and simple argument for the blueward extent of the technical concept, i.e. inclusion of the 308 nm OH line. In addition, sheer throughput is the main requirement. An attempted observation of a main-belt comet with X-Shooter produced only upper limits to OH emission (Snodgrass et al. 2017b), so the throughput needs to be significantly better at this wavelength than that for the bluest order of X-Shooter. In contrast to the stellar cases above, observations at $R$ of a few thousand are sufficient (although higher-resolution observations could be rebinned spectrally to regain most of the potential sensitivity).

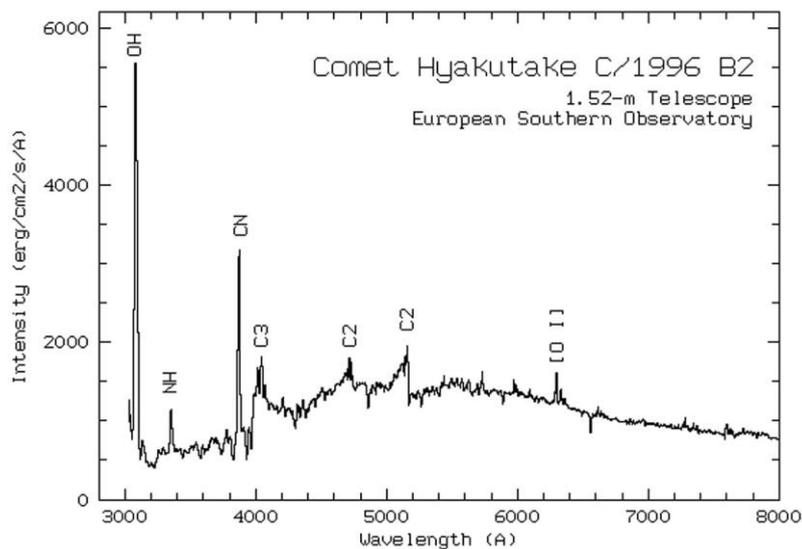

Fig. 3: OH emission at 308 nm is a potentially powerful probe of water outgassing from main-belt comets, but we need better sensitivity than available with present facilities (see Snodgrass et al. 2017a). Image credit: ESO.

## 3. EXTRAGALACTIC SCIENCE

Near-UV spectroscopy will also provide unique extragalactic observations. For example, in studies of the circumgalactic medium (CGM) of distant galaxies, where gas in the halo of a foreground galaxy imprints absorption lines on a background source, and in measuring the contribution of different types of galaxies/AGN to the cosmic UV background.

### 3.1. Missing baryonic mass in the high-redshift CGM

A vibrant area of research is the so-called 'missing baryon' problem, where the detected baryonic mass compared to the universal fraction suggests that we are missing more than 60% in galaxies outside of clusters, rising to ≥90% for gas-dominated galaxies and dwarfs (e.g. McGaugh et al. 2010). Recent studies with *HST* of the CGM in galaxies at $z$~0.2 concluded that diffuse gas in their halos could account for at least half of the previously missing baryonic mass (Werk et al. 2014). Although large uncertainties remain, a substantial fraction of mass appears to be contained in diffuse halo components and there is considerable interest in whether this was also the case at earlier times.

One of the cases developed for with the ELT-MOSAIC instrument (Puech et al. these proceedings) is to use pairs of Lyman-break galaxies out to $z$~3 to probe the CGM in more distant systems. Observations in the ground-UV can open-up an exciting gap in this field (cf. *HST* and ELT), providing access to galaxies at $z$ = 1.5 to 2 (see Fig. 3), immediately after the era of peak star-formation in the Universe (e.g. Madau & Dickinson, 2014). The relatively low density of CGM clouds at these redshifts also means that, with observation at medium resolution ($R$~20,000), the impact of the Lyα forest on the measurement of the CGM metal lines can be minimised (which is not the case at larger redshifts).

The new Keck Cosmic Web Imager (KCWI) instrument (Morrisey et al. these proceedings) has partly been designed to address this topic and will certainly make significant contributions, but its blueward coverage stops at 350 nm. Observations shortward of this with the VLT will enable, e.g., observations of the important OVI line (used to trace the warm-hot gas) to lower redshifts. A significant improvement in efficiency cf. UVES will open-up background sources 1-2 magnitudes fainter than the QSOs used at present. This will also provide new opportunities to investigate metal abundances in the IGM; current studies are limited to the brightest QSOs (e.g. D'Odorico et al. 2016) and greater sensitivity would open-up new sightlines to fainter galaxies.

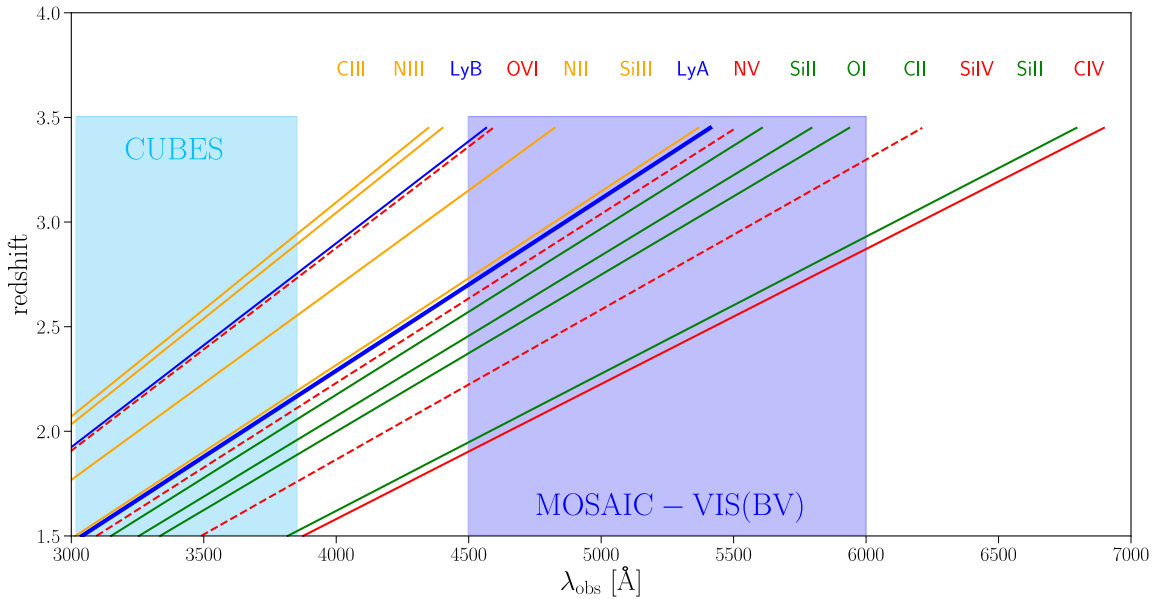

Fig. 3: Lines available in the ground-UV for studies of the circumgalactic medium of galaxies at 1.5<$z$<2. They span a range of ionisation states (green = low, yellow = intermediate, red = high) to study the relative fractions of cold vs. warm vs. hot gas, at redshifts where the contamination of key ions such as OVI by the Lyα forest is less severe. These observations will neatly complement future observations observations with the blue-visible waveband of ELT-MOSAIC, which is targeting the CGM in distant galaxy halos at $z$~3 (Puech et al. & Morris et al. these proceedings).

### 3.2. Cosmic UV background

Cosmic reionisation is a major focus of cosmology, but, after almost 40 years (Sargent et al, 1980), the sources driving it and keeping the Universe ionised are still not understood. Galaxies are commonly thought to be able to produce the bulk of the UV emissivity at high redshift (e.g., Robertson et al. 2015) but the AGN population is also proposed as a relevant or dominant contributor (Giallongo et al. 2015; but also see Fontanot et al. 2012 and Haardt & Salvaterra 2015, for different views) and more exotic possibilities such as decaying particles cannot be excluded (e.g. Poulin et al. 2015).

Quasars are known to be efficient in producing UV photons with the fraction ($F_{esc}$) of ionizing photons able to escape to the IGM of close to 100% (Cristiani et al. 2016, Grazian et al. 2018). However, there are doubts whether at $z>4$ their volume density, still poorly known at intermediate/low luminosities, can ensure the required ionisation rate. Galaxies are much more numerous, but $F_{esc}$ has large uncertainties, with only a few reliable detections of galaxies 'leaking' significant Lyman continuum (LyC) flux so far, namely: Ion2 at $z=3.218$ (Vanzella et al. 2015), Q1549-C25 at $z=3.15$ (Shapley et al. 2016), A2218-Flanking at $z \approx 2.5$ (Bian et al. 2017), Ion3 at $z=3.999$ (see Fig. 4; Vanzella et al. 2018).

The percentage of H-ionising photons (produced by massive stars) able to escape the ISM of a galaxy is the combination of multiple complex phenomena. It probably depends on the star-formation intensity (expected to be larger at high redshift), SNe explosions, the depth of galactic potential wells, the geometry of the HII regions, and the dynamics of the absorbing gas. To be detected by an observer at $z=0$ the emerging photons have to survive IGM absorption which increases strongly with redshift and is significantly variable between sightlines (it is dominated by the incidence of relatively small numbers of intervening HI systems with typical column densities of $\sim 10^{16}$-$10^{18}$ cm$^{-2}$). Given these often competing factors, the ideal redshift range for ground-based observations of the LyC of a galaxy (at $\lambda_{obs} < 912(1 + z_{gal})$) is $2.3 < z < 3.5$, i.e. from ~410 nm down to the atmospheric cutoff.

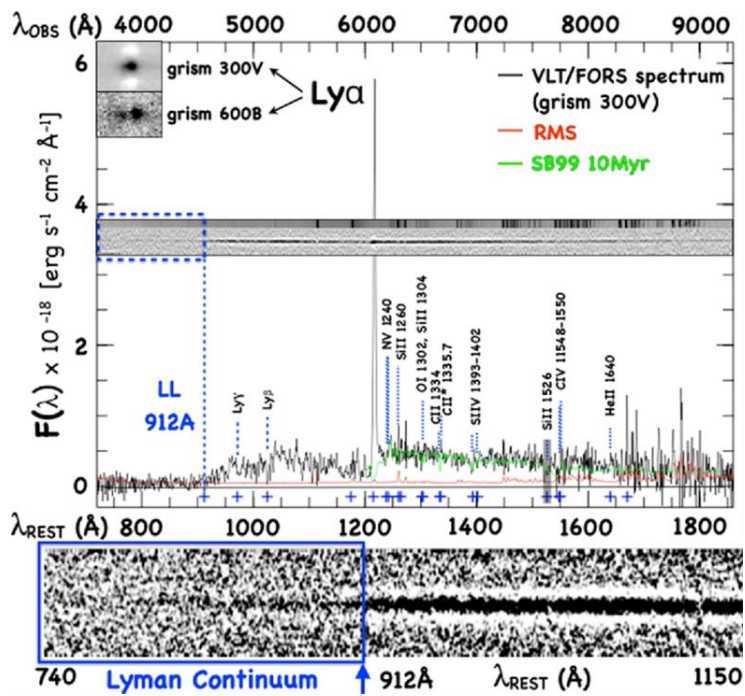

Fig. 4: FORS 300V grism spectrum of Ion 3, with the most relevant UV lines identified (Vanzella et al. 2018). The insets at the top left show the Lyα spectra obtained with the 300V and 600B grisms (resolutions of $\delta v \simeq 580$ and 300 km s$^{-1}$, respectively). The two-dimensional signal-to-noise and sky spectra are shown in the middle of the figure, in which the wavelength coverage up to 9300 Å is well sampled, despite the fringing pattern at $\lambda > 7800$ Å (typical of the e2v blue-optimized CCD). In the bottom panel the emerging LyC (with S/N>10) is clearly detected.

For a significant dynamical range in parameter space and to minimize the effects of fluctuations in the intervening IGM opacity, there are strong and somewhat competing requirements for ground-based observations at high redshift:
- To maximise the sensitivity of measurements of individual sources, i.e. where significant LyC flux is detected without stacking, as close as possible (at relatively high spectral resolution) to the Lyman limit of each;
- To minimise the systematics affecting the measurement of individual sources (e.g., background subtraction, contamination by other sources) and those hampering the statistical assessment of the intervening IGM absorption (including pointings in different parts of the sky, to avoid the effects of large-scale structures, and an accurate characterisation of the IGM absorption statistics).

A large part of recent observational efforts has been dedicated to imaging surveys, which have an obvious multiplex advantage. Intermediate- and narrow-band filters can be tuned to select the emission just below the rest-frame Lyman limit at a redshift of interest (e.g. at the redshift of a protocluster, Mostardi et al. 2013). Alternatively, extremely deep broad-band imaging is used to search for LyC emission from galaxies with an appropriate (known) spectroscopic redshift (i.e. those for which the rest-frame Lyman limit lies immediately redward of the observing band, e.g. Grazian et al. 2017). In spite of the multiplex advantage, putative detections (and the quantification of non-detections) require both follow-up spectroscopy and high-resolution (typically *HST*) imaging (e.g. Vanzella et al. 2010). Most recently, the Keck Lyman Continuum Spectroscopic Survey (KLCS, Steidel et al. 2018) has detected LyC in 15 of 124 target galaxies (for galaxies at redshifts of $2.9 < z < 3.3$) from total integrations of 8-13 hrs with Keck-LRIS.

These direct spectroscopic detections are valuable but the challenge for the future is to go beyond them to characterise which particular properties of galaxies determine their propensity to 'leak' LyC radiation. Recent progress has been made at low redshift ($z$~0.3-0.5) via observations of compact, low-mass, star-forming galaxies with stellar masses, metallicities and star-formation rates as similar as possible to those of high-$z$ Lyα emitting galaxies (Izotov et al. 2018). The use of diagnostics such as a high $O_{32}$ line-ratio (i.e. [OIII]5007/[OII]3727 > ~5) to indicate they may contain density-bounded HII regions and be potential LyC leakers has been shown to be useful. However, it is difficult to find reliable analogues of the high-$z$ ionising systems in the local universe, and the possibility of observations at low-$z$ in the near future is tied to continued operations of the *HST*. In this context, observations with a far-blue, efficient spectrograph on the VLT will be particularly powerful. In addition to ensuring high throughput, such an instrument needs to be sky-limited at intermediate/high resolution (needed for disentangling in cases of contamination, e.g. Fig. 5) and with a relatively long slit for accurate background subtraction.

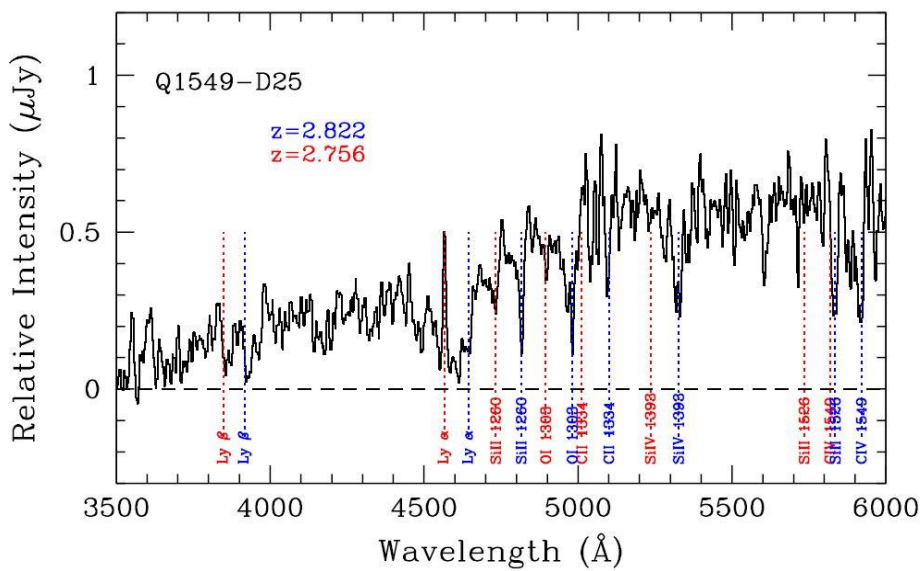

Fig. 5: Example candidate LyC leaker identified as a spectroscopic blend of two redshifts within the 1-D extraction footprint of the primary target. Line identifications are marked as indicated for the two redshifts (from Steidel et al. 2018).

## 4. SUMMARY & NEXT STEPS

Alongside an overarching goal of maximising the instrument throughput to best exploit the VLT in this wavelength domain, the two key requirements for the Phase A conceptual design were a spectral resolving power of $R \geq 20,000$ spanning 302-380 nm, with extension to 400 nm as a goal (ensuring good overlap with ESPRESSO). After revisiting the scientific case these are still valid. As noted in the introduction, we anticipate demand from the community to address more topics than those described here. For instance, observations of planetary nebulae, close binaries, symbiotic stars, and several AGN cases were discussed by Barbuy et al. (2014), and the detailed Phase A case included a significant section on interstellar studies in the local Universe. Moreover, we expect significant interest in the context of transient follow-up, particularly given new facilities such as MeerLICHT and BlackGEM (see Groot et al. papers from SPIE conf. #10700 in parallel to this meeting) and the start of LSST operations in the early 2020s. We will build on the summaries here to develop a revised and expanded science case.

We are now planning the technical work to advance the instrument concept beyond that presented by Barbuy et al. (2014). The dispersion grating is a vital ingredient in delivering the high efficiency desired – a prototype fused-Si transmission grating has been manufactured by Fraunhofer IOF and will be characterised in the coming year. Another aspect of the technical design is the use of an image slicer to minimise slit losses while preserving spectral resolution – we will revisit the science trades for the slicer design before optimising the overall opto-mechanical concept.

To conclude, in the context of current plans for Paranal and the future instrument suite of the ELT, a high-efficiency spectrograph working in the ground-UV domain will open-up unique discovery space for the VLT for years to come. We also comment that it will be a relatively modest instrument in terms of the hardware costs and staff effort compared to recent second-generation VLT facilities.

## ACKNOWLEDGEMENTS

We acknowledge support from the Global Challenges Research Fund (GCRF) from UK Research and Innovation. We gratefully acknowledge Olivier Hainault's contribution of the original comet case in the past Phase A study, and Hans Dekker and Bernard Delabre for their recent inputs to the technical design. R.S. acknowledges support from the Polish Ministry of Science and Higher Education.